\documentstyle[12pt]{article}
\begin{document}

\title{The Cosmological Constant, False Vacua, and Axions}

\author{{\bf S.M. Barr} \\ and \\ {\bf D. Seckel} \\
Bartol Research Institute \\ University of Delaware \\
Newark, DE 19716}

\date{BA-01-xx}
\maketitle

\begin{abstract}

It is suggested that the true ground state of the world has
exactly vanishing vacuum energy and that the cosmological constant
that seems to have been observed is due to our region of the
universe being stuck in a false vacuum, whose energy is split
from the true vacuum by non-renormalizable operators that are
suppressed by powers of the Planck scale. It is shown that
conventional invisible axion models typically have the features 
needed to realize this possibility. In invisible axion models
the same field and the same potential can explain both the cosmological
constant (or dark energy) and the dark matter. It is also shown that
the idea can be realized in non-axion models, an example of which
is given having $\Lambda = M_W^7/M_{Pl}^3$, which accords well with the
observed value.

\end{abstract}

\newpage

\section{Introduction}

From the basic facts of cosmology one knows that the cosmological
constant must be smaller than or of order $10^{-123} M_{Pl}^4$.
($M_{Pl} \equiv G_N^{-1/2} \cong 1.2 \times 10^{19}$ GeV.)
This follows from the value of the Hubble parameter $H \sim
(10^{10} {\rm yrs})^{-1} \sim 10^{-42} {\rm GeV}$ and the 
fact that in a cosmological constant dominated universe 
$H^2 = \frac{8 \pi}{3} (\Lambda/M_{Pl}^2)$.
Such a fantastically small limit on $\Lambda$ in gravitational units
naturally suggested to theorists that in fact it was exactly zero.
In most discussions of cosmology $\Lambda$ was indeed simply set to zero.
While no principle or mechanism was known that would explain why 
$\Lambda$ vanished, it was expected that eventually one would come 
to light.

For this reason, the recent observations \cite{sn} on type-I supernovas that
seem to indicate a non-zero value of $\Lambda$ came as quite a shock.
The observations, if indeed they are to be
explained by a cosmological constant, correspond to a value of $\Lambda 
\cong (2 \times 10^{-3} {\rm eV})^4 \cong 10^{-123} M_{Pl}^4$.
This is about 70\% of the critical density. 
Observations are consistent
with a spatially flat universe, which would imply that
all matter (ordinary plus dark) add up to the remaining 30\%
of critical density. These figures compound the mystery: why should
the cosmological constant be so near in value to the matter density that
the universe happens to have at the {\it present} time?  
This flies in the face of
the hallowed ``Cosmological Principle" \cite{cosprinc} that there is nothing
special about the cosmological time or place in which we live.

The original Cosmological Constant Problem \cite{ccp} was the problem of
finding a principle or mechanism that made $\Lambda =0$, at
least approximately. We may call that the Old Cosmological Constant
Problem. We now have in addition what may be called the New 
Cosmological Constant Problem, which is to explain why $\Lambda$ is 
not {\it exactly} zero,
and more specifically to explain the origin of the very small
number $10^{-123}$ in terms of fundamental physics. Finally,
there is the problem of accounting for the similarity of $\Lambda$
and $(\rho_{matter})_{now}$. This is sometimes called the Cosmological
Coincidence Problem \cite{ahkm}.

There are several promising approaches to solving these 
problems. It is worth mentioning four of them.

{\bf (1) The Anthropic Principle.}
One possibility is that the value of $\Lambda$ is explained by the
so-called Weak Anthropic Principle \cite{ap}. Many people are put off by this
name, but the name is misleading. The Weak Anthropic Principle is
neither weak, nor anthropic, nor a principle. It is really
just the old idea of observer selection or observer bias.
If the universe has many domains having different values of the
cosmological constant, then, naturally, observations of the cosmological
constant can only be made in those domains where it has
a value compatible with the existence of observers. Weinberg has 
used the assumption that observers are unlikely to exist where galaxies 
cannot form to set an ``anthropic" upper limit on observable values of 
$\Lambda$ \cite{weinberg}.This limit comes out to be 
of the same order as the present value of $\rho_{matter}$.
There is no reason, on the other hand, why values of $\Lambda$  
much {\it less} than this anthropic limit should be preferred. Therefore, 
if $\Lambda$ does vary among domains, 
one should expect to observe a value of $\Lambda$ that is of order 
$\rho_{matter}$ today. The beauty of this approach is that it can solve
all three cosmological constant problems at once.

{\bf (2) Quintessence.}
The idea of quintessence \cite{wetterich} \cite{peebles} 
\cite{quint} is that there is some slowly rolling
field, call it $\phi$, whose energy accounts for the apparent
cosmological constant. The quintessence idea makes no attempt to
resolve the Old Cosmological Constant Problem. It simply assumes that
at the minimum of the quintessence field's potential (which may be at
$\phi = \infty$) the
cosmological constant vanishes: $V(\phi_{min}) = \Lambda_0 = 0$. 
It is assumed, however, that the field $\phi$ started away from its
minimum and is still slowly approaching it. In some models the
quintessence energy roughly ``tracks" the matter density \cite{track}, thus
explaining why $\Lambda_{eff} \sim \rho_{matter}$ today. 

{\bf (3) Dynamical Relaxation.}
The idea of dynamical relaxation is superficially similar to the
quintessence idea, but has the more ambitious goal of resolving the
Old Cosmological Constant Problem. As in quintessence models, there
is a field (again call it $\phi$) that is slowly rolling and causing
the effective cosmological constant to vary with time. (Indeed, this
possibility seems to have been considered first in the context of 
dynamical relaxation of $\Lambda$ \cite{barr}.) However, unlike 
ordinary quintessence models, dynamical relaxation models
do not assume that the potential $V(\phi)$ has a minimum at
which the vacuum energy vanishes. For to assume that 
is equivalent to simply setting $\Lambda$ to zero, and this ``fine-tuning"
is what the dynamical relaxation idea seeks to avoid. Rather, what is
assumed is that there is a dynamical feedback mechanism which
in some indirect way
``tells" the field $\phi$ when $\Lambda_{eff}$ is approaching zero, 
and stops or slows the rolling of $\phi$
even though it is not at a minimum of its
potential. For example, suppose that the action of $\phi$ couples it to
the scalar curvature of spacetime, $R$. The scalar curvature, in turn,
``knows" about $\Lambda_{eff}$ through the trace of Einstein's
equation. Thus, as $\Lambda_{eff}$ approaches zero through the rolling of
$\phi$, $R$ also approaches zero, and it is possible, given the right
form of $S(\phi, R)$, that the rolling of $\phi$ will be braked.

It is actually possible to construct models where this happens, and
where $\Lambda_{eff}$ asymptotically approaches zero even though
the potential has {\it not} been fine-tuned to make $V(\phi_{min}) 
= 0$ \cite{barr} \cite{hw}.
In \cite{barr}, the scalar curvature $R$ acts as the go-between in
the dynamical feedback loop. In \cite{hw}, a Brans-Dicke type scalar
field plays the same role. An interesting feature of the models
constructed in Ref. \cite{barr} was that $\Lambda_{eff}$ approached zero
at roughly the same rate as did the matter density.
The reason for this is not entirely clear, but it may be a necessary
feature of such feedback models. The point is that if $\Lambda_{eff}$
falls far below $\rho_{matter}$, the effect of $\Lambda_{eff}$ on
the rolling of the scalar field may become negligible compared to the effect
of $\rho_{matter}$, which would break the feedback loop. 
While, on the other hand, if $\Lambda_{eff}$ falls off
more slowly than $\rho_{matter}$, it would come to dominate over 
$\rho_{matter}$ early in the
universe, leading to an inflationary phase with no exit. In any event, 
if it is indeed true that $\Lambda_{eff}$ must
relax in such a way as to stay roughly of order $\rho_{matter}$, it
would explain the currently observed value of the cosmological ``constant".
Such a scenario has the possibility of solving all three problems
relating to the cosmological constant.

{\bf (4) False vacuum.}
The fourth possibility is similar to the quintessence idea
in that it is assumed that in the true ground state the vacuum energy
vanishes exactly because of some (as yet unknown) fundamental symmetry 
or mechanism. However, unlike quintessence, it is not assumed that
the universe is slowly approaching its true ground state; rather,
it is assumed that the universe is stuck in a false vacuum. 
The idea is that the resulting $\Lambda_{eff}$ is extremely
small because the splitting between the false and true vacua
arises from very high order effects \cite{gc} or from non-perturbative 
\cite{goldberg} effects.
This is the approach we shall pursue in this paper.

This idea can naturally solve the New Cosmological Constant Problem.
But how could it explain the ``Cosmic Coincidence"? Unlike the
other approaches, here there is no obvious way that the density of matter
comes into the calculation of $\Lambda_{eff}$. An answer is
suggested \cite{ahkm} by a well-know piece of ``numerology". If one writes
the value of $\Lambda$ 
suggested by recent observations as $\Lambda = M^8/M_{Pl}^4$, then 
one finds that $M \approx 5$ TeV, which is
close to the Weak interaction scale. Alternatively,
if one writes it as $\Lambda = M^7/M_{Pl}^3$, then $M \approx 35$ GeV,
also quite close to the Weak Scale. This suggests the possibility that
the false vacuum state we are in may be split from the true vacuum
by higher-dimension
operators that go as $\Phi^8/M_{Pl}^4$ or $\Phi^7/M_{Pl}^3$,
where $\Phi$ is a field
associated in some way with Weak interaction breaking. 

What is appealing about this idea is that it is similar in spirit
to the way the values of many other small parameters in particle
physics are explained. First, the smallness of the parameter
is explained as being the result of some symmetry principle. Second, 
the departure of the parameter
from zero is explained as being due to a small breaking of the symmetry.
And finally, the actual size of the parameter is explained by
relating the symmetry breaking to some known dynamical scale.
In the present case, we do not know the symmetry principle
(or other principle) that makes $\Lambda$ vanish in the absence
of breaking. But we can still try to relate the breaking that generates
the observed $\Lambda$ to some known physics such as the Weak scale.
This is what we shall attempt to do.

The idea can be
simply illustrated with a toy example. Suppose that $\Phi$ has the
following potential:

\begin{equation}
V(\Phi) = \lambda(\Phi^{\dag} \Phi - f^2)^2
- [a \Phi^6/M_{Pl}^2 + b e^{i \beta} \Phi^8/M_{Pl}^4 + H.c.],
\end{equation}

\noindent
where $f \sim M_W$ and $a,b \sim 1$. Any phase in $a$ can be absorbed by
a redefinition of $\Phi$.
With $\Phi = (f + \tilde{\rho}) e^{i \theta}$,
the potential for $\theta$ is 

\begin{equation}
V(\theta) = - 2 a (f^6/M_{Pl}^2) \cos (6\theta) 
- 2 b (f^8/M_{Pl}^4) \cos (8 \theta + \beta).
\end{equation}

The first term has six degenerate minima at $\theta = \frac{1}{6}
(2 \pi N)$. These are split by the second term, which 
contributes $\Delta E_0 = \Delta E_3 = -2 b (f^8/M_{Pl}^4)
\cos \beta$, $\Delta E_1 = \Delta E_4 = - 2 b (f^8/M_{Pl}^4)
\cos (\beta + 2\pi/3)$, and 
$\Delta E_2 = \Delta E_5 = - 2 b (f^8/M_{Pl}^4)$ $\cos(\beta + 4\pi/3)$. 
Thus, there are in this example
false vacua that are split from the true vacuum by an energy
density of the desired order to explain the observed
cosmological constant. Unfortunately,
there are several problems with the toy model just described.

(1) Although in this toy model
there exist false vacuum states, the universe
would not end up in the present epoch sitting in one of them.
What happens, rather, is the following. At a temperature 
$T \sim f \sim M_W$ the field $\Phi$ develops an expectation
value, breaking the global $U(1)$ symmetry of the
renormalizable part of the potential. At this point cosmic strings
appear. These strings undergo a complicated evolution which is such that
at any given time there is of order one ``infinite string"
(i.e. string longer than the horizon length) per horizon volume.
When the thermal energy density drops to a value comparable to
the barrier caused by the term $a \Phi^6/M_{Pl}^2$, domain walls 
form which have mass/area $\sigma_w \sim \sqrt{a} f^4/M_{Pl}$.
The vacua separated by these walls have a pressure difference
of $P \sim b f^8/M_{Pl}^4$. This pressure difference will tend to
squeeze out the false vacua on a time scale of order
$\sigma_w/P \sim b a^{-1/2} M_{Pl}^3/f^4$, which is comparable
to the present age of the universe. 
One sees that, depending on the values of the 
dimensionless parameters $a$ and $b$, either the false vacuum will 
be gone by now, or the system of walls will still be in existence.
Neither situation corresponds to a non-zero cosmological constant
throughout the observable universe. 

(2) The second problem is to explain why the global $U(1)$
symmetry that gives the desired approximate degeneracy of the minima
is not broken by lower dimension operators such as $m^2 \Phi^2 + H.c.$.
If it is, then these terms dominate the potential for $\theta$ and force
it to take one of the values $\theta = \pi N - \theta_m$, 
which are not split in energy by the higher-dimension terms. Similarly,
if there is no $\Phi^2$ term, but there is a $\lambda e^{i \gamma} 
\Phi^4 + H.c.$
term, the phase will be forced to take one of the values
$\theta = \pi N/2 - \gamma/4$, which are not split 
in energy by the $\Phi^8$ term, as desired, but are split by the
$\Phi^6$ term.

In other words, there must be a local symmetry that forbids  
low order terms that break the global $U(1)$ but allows
higher order terms, as shown in Eq. (1).

(3) Finally, the model as it stands does not relate the field
$\Phi$ to the breaking of the Weak interactions, and so does not
explain the connection between the parameter $f$ in Eq. (1) 
and the Weak scale. Certainly $\Phi$ is not just the Standard Model
doublet, as then Weak isospin does not allow the higher-dimension
terms in Eq. (1). 

In Sections 2 and 3 of this paper we shall look at two types of model that
seek to overcome these problems and implement the basic idea
illustrated by the toy model in Eq. (1). The first type of model is based
on familiar physics, namely the invisible axion. We point out that
simple invisible axion models of the kind that have long been studied
can quite naturally realize the idea that the observed cosmological
constant is false vacuum energy. In such a scenario the cosmological
constant is not given in terms of the Weak scale and Planck scale,
but rather in terms of the Peccei-Quinn scale $f_a$ and the Planck scale.
If $f_a$ is in the window between $10^{10}$ GeV and $10^{12}$ GeV,
as suggested by the requirement of solving the ``axion energy problem", 
then one can invoke inflation to explain how our domain of the universe
ended up in one of the false vacua. As a bonus, the energy in the coherent 
axion oscillations
can be the dark matter. One has thus the beautiful possibility that
the same field and the same potential 
solve the ``dark matter" and ``dark energy" problems (and the strong CP
problem!) at the same time.

The second type of model we discuss attempts to explain the cosmological
constant directly in terms of the Weak scale and and the Planck scale
using a potential similar to that in Eq. (1). In order to explain
how the universe ended up in a false vacuum, inflation is invoked.
This entails the introduction of additional fields 
which have superlarge expectation values. These large
expectation values allow certain complex phases to be laid down at
a time prior to inflation. 

Both of these models invoke inflation to explain
special initial conditions. Inflation would also induce fluctuations in
the equation of state and therefore in the microwave background.
In Section 4 we consider these fluctuations and determine the
constraints they place on inflation. In the same section we also discuss 
some alternatives to inflation for explaining how the universe ended in a 
false vacuum.

\section{Invisible Axion Models}

The axion solution to the strong CP 
problem is based on the existence of the global $U(1)$ 
Peccei-Quinn symmetry \cite{pq}. However, there are various reasons to
believe that quantum gravity does not respect global symmetries, 
and consequently, as was pointed out a long time ago, 
Planck-scale physics should induce explicit breaking of the 
Peccei-Quinn symmetry \cite{holman} \cite{kam} \cite{bs}. 
This would destroy the Peccei-Quinn mechanism
unless some symmetry that was local, and thus respected by quantum gravity,
prevented explicit Peccei-Quinn-breaking terms from arising
up to a sufficiently
high order in $1/M_{Pl}$. This is one of the challenges in constructing
viable axion models, but it also provides a natural way of explaining
a very small cosmological constant. Specifically, 
these same higher-dimension
explicit Peccei-Quinn-breaking terms that ought to exist because of
Planck-scale physics can lift the degeneracy among
the several vacuum states that one typically finds in axion models.

Consider a typical invisible axion model,
in which the Peccei-Quinn symmetry is spontaneously broken by a field
$\Omega$ at a scale $f_a$ that is between $10^{10}$ and $10^{12}$ GeV. 
Let the potential for $\Omega$ be

\begin{equation}
V(\Omega) = - \lambda( \Omega^{\dag} \Omega - f_a^2)^2 
+ [b e^{i \beta} \Omega^n/M_{Pl}^{n-4} + H.c.].
\end{equation}

\noindent
We have added to the usual ``Mexican hat" potential that is invariant
under the Peccei-Quinn symmetry a non-renormalizable term, assumed to
come from Planck-scale physics, which explicitly violates $U(1)_{PQ}$.
One can write $\Omega(x)  = (f_a + \tilde{\Omega}(x)) e^{i a(x)/f_a}$, 
where $a(x)$ is the axion field. The axion field gets mass from
two sources: (1) QCD instanton effects and (2) the non-renormalizable
operators in Eq. (3).
Suppose that $\Omega$ has a Peccei-Quinn charge of $q$ and that
QCD instantons break $U(1)_{PQ}$ down to $Z_N$. Then, the potential for
the axion field $a(x) \equiv \theta f_a$ is given by

\begin{equation}
V(\theta) =  - g f_{\pi}^4 \cos (N \theta) + 2 b (f_a^n/M_{Pl}^{n-4})
\cos (n q \theta+ \beta),
\end{equation}

\noindent
where $g \sim 1$. The first term in Eq. (4) is the QCD-instanton-generated 
potential. Assume that $\theta$ is small;
then one may write

\begin{equation}
V(\theta) \cong \frac{1}{2} g f_{\pi}^4 N^2 \theta^2 
+ 2 b (f_a^n/M_{Pl}^{n-4})
(\cos \beta - (nq \theta) \sin \beta).
\end{equation}

\noindent
This gives $\theta_{min} \equiv \overline{\theta} 
\cong \frac{2 b n q \sin \beta}{g N^2}
\left( \frac{f_a^n}{f_{\pi}^4 M_{Pl}^{n-4}} \right)$. Demanding that
$\overline{\theta} \leq 10^{-9}$, to solve the strong CP problem, one
has that $n \geq 13$ for $f_a = 10^{12}$ GeV, and $n \geq 10$ for 
$f_a = 10^{10}$ GeV. The question of how local symmetries can
prevent explicit Peccei-Quinn-breaking operators from arising
up to such very high orders has been shown in the literature to
be answerable in various ways \cite{holman} \cite{kam} \cite{bs} \cite{bb}. 
We shall return to this 
question shortly. 

What value of $n$ allows us to account for the cosmological constant?
We see from Eq. (4) that the instanton-generated potential for the
axion field has $N$ degenerate minima. However, this degeneracy
is lifted by the second term in Eq. (4) (unless it happens that
$nq$ is a multiple of $N$) by an amount of order 
$f_a^n/M_{Pl}^{n-4}$. Thus, assuming that the lowest of these minima ---
the true vacuum --- has vanishing cosmological constant by some
as-yet-not-understood mechanism, and that our region of the universe is
in one of the
other minima, we would observe a cosmological constant
$\Lambda \approx f_a^n/M_{Pl}^{n-4}$. If we set this equal to the
observed cosmological constant, then $n \cong 17$ for $f_a = 10^{12}$ GeV,
and $n \cong 13$ for $f_a = 10^{10}$ GeV. Thus the values of $n$
required to explain the cosmological constant are safely
larger than the lower limit coming from a satisafactory solution
to the strong CP problem.

To see how local symmetries can protect the Peccei-Quinn symmetry up
to sufficiently high order in $1/M_{Pl}$, and to verify that the
Planck-scale operators really can lift the degeneracy of the
instanton-generated potential to produce a cosmological
constant, we present a simple toy model constructed along the 
lines suggested in \cite{bs}

Consider a model whose gauge symmetry is
$SU(3)_c \times SU(2)_L \times U(1)_Y \times U(1)'$. Let there be
$pN$ flavors of left-handed anti-quarks with $U(1)'$ charge $-q$, denoted
$\overline{Q}_{-q}$; $qN$ flavors of left-handed anti-quarks with $U(1)'$
charge $+p$, denoted $\overline{Q}_p$; and $(p+q)N$ flavors of left-handed
quarks with $U(1)'$ charge equal to zero, denoted $Q_0$.
This set of quarks obviously has no $SU(3)_c^2 \times U(1)'$ anomaly, while 
the $U(1)^{\prime 3}$ anomaly can be cancelled by
fermions that have no color. Let the quarks obtain mass from two
scalar fields that have $U(1)'$ charges $p$ and $q$, denoted 
$\Omega_p$ and $\Omega_q$. 

\begin{equation}
{\cal L}_{mass} = \overline{Q}_p Q_0 \Omega^*_p 
+ \overline{Q}_{-q} Q_0 \Omega_q.
\end{equation}

\noindent
We suppress the flavor indices of the quarks.
The fields $\Omega_p$ and $\Omega_q$ acquire vacuum
expectation values of order $f_a$. If we assume that $p$ and $q$
are relatively prime, then the lowest-dimension operator
that knows about the relative phase of $\Omega_p$ and $\Omega_q$
is $O_{(p+q)} \equiv
(\Omega_p)^q (\Omega_q^*)^p/M_{Pl}^{p+q-4}$. If such 
higher-dimension operators are neglected, the model clearly has
a $U(1) \times U(1)$ symmetry corresponding to independently
rotating the phases of $\Omega_p$ and $\Omega_q$. One combination
of these $U(1)$ symmetries is just the local symmetry $U(1)'$.
The other is an anomalous global $U(1)$ symmetry, i.e. a
Peccei-Quinn symmetry. One sees that the local $U(1)'$ prevents
any explicit Peccei-Quinn-breaking operator up to order $n = p + q$.

Instanton effects
break the Peccei-Quinn symmetry down to $Z_N$, as can be shown in the
following way. Since $p$ and $q$ are relatively prime, there
exist integers $a$ and $b$ such that $pa + qb = 1$. Consider
a rotation of the fields that takes $\Omega_p \rightarrow
e^{2 \pi b/N} \Omega_p$ and $\Omega_q \rightarrow e^{- 2 \pi a/N}
\Omega_q$. It takes $N$ such rotations to bring the fields back to
their original values. Thus, this rotation generates a $Z_N$ transformation 
of the fields. From Eq. (6) and the fact that there are $pN$ 
flavors of $\overline{Q}_{-q}$ and $qN$ flavors of $\overline{Q}_p$ one sees
that this rotation changes the QCD CP angle by $\Delta \overline{\theta}
= (2 \pi a/N)(p N) + (2 \pi b/N)(q N) = 2 \pi$. Thus the instanton
potential is left invariant by this $Z_N$ rotation. On the other
hand, the higher-dimension operator $O_{(p + q)}$ is changed by a phase 
$(2 \pi a/N) p + (2 \pi b/N) q = 2 \pi /N$. Thus, the lowest-dimension
Peccei-Quinn-breaking operator allowed by $U(1)'$ does indeed lift the 
$N$-fold degeneracy of the instanton potential. 
This example gives a KSVZ type of invisible axion \cite{ksvz}. It
is a simple matter to construct a model of the cosmological constant
in a DFSZ type of axion model \cite{dfsz} using the ideas in \cite{bb}.

Since the Peccei-Quinn symmetry is broken at a scale $f_a$ that
is between $10^{10}$ and $10^{12}$ GeV, one can assume that an inflation
occurs after that transition and gets rid of axion strings and
axion domain walls. If so, then our whole observable universe
would be in a region with an essentially constant axion field value 
and consequently would all fall into the same minimum of the instanton
potential. In that way we could end up in one of the false vacua.
There would also be coherent axion fluctuations about that minimum, 
which could act as the cold dark matter if $f_a \sim 10^{12}$ GeV.
In this case, both the ``dark matter" and the ``dark energy" would
have their origin in the energy of the same field.

What is appealing about this possibility is that it does not
require that any new physics be introduced just for the purpose of
explaining the magnitude of the cosmological constant. The axion was 
introduced to solve the strong CP problem, and it is a typical feature of 
axion models that they have both degenerate minima and higher-dimension
operators that lift this degeneracy by an amount that is of high order
in $1/M_{Pl}$.

\section{Models where $\Lambda$ arises from electroweak breaking.} 

Consider a supersymmetric model in which the superpotential
contains the following terms

\begin{equation}
W \supset -  a (H_u H_d) X_1^2/M_{Pl}
-  b (H_u H_d)^2 Y_4/ M_{Pl}^2 
- c (H_u H_d)^3 Y_6/ M_{Pl}^4.
\end{equation}

\noindent
Any phases in the coefficients $a$, $b$, and $c$ can be absorbed into
redefined superfields $H_u$, $H_d$, $X_1$, $Y_4$ and $Y_6$.
$H_u$ and $H_d$ are just the Higgs doublet superfields of the MSSM.
$X_1$ is a superfield whose scalar component is assumed to get a
superlarge vacuum expectation value, which we shall denote $f$. The 
subscript ``1" refers to its charge under a local abelian symmetry 
$U(1)'$. Clearly, the product $H_u H_d$, which is 
a singlet under the Standard Model gauge group, has a $U(1)'$ charge of $-2$. 
Note that the first term in Eq. (7) generates the $\mu$ parameter
of the MSSM. Demanding that $\mu \sim M_W$, implies that
$f \sim a^{-1/2} \sqrt{M_W M_{Pl}}$. By $M_W$ we mean the Weak scale in some 
rough sense, rather than the mass of the $W$ boson. We will take $M_W
\sim 10^2$ GeV. 
The superfields $Y_4$ and $Y_6$ have $U(1)'$ charges of
4 and 6 respectively, and are assumed to have expectation values
of order $M_{Pl}$.

Denote the phases of $(H_u H_d)$, $X_1$, $Y_4$, and $Y_6$
by $\theta_h$, $\theta_1$, $\theta_4$, and $\theta_6$ respectively.
One linear combination of these is the gauge
degree of freedom that is eaten by the $U(1)'$ gauge boson. The
other three linear combinations are physical phases.
It is possible to couple the scalars to colored fields is such a way
that the global symmetries corresponding to the rotation of these
phases has no QCD anomaly. Therefore, in the potentials to be
considered below there is no need for an instanton contribution.
After supersymmetry breaking the terms in Eq. (7) lead to ``A terms"
which have the same form. (We assume high-scale supersymmetry breaking.)
This leads to a potential for the phases of the form

\begin{equation}
\begin{array}{cl}
V(\theta_h, \theta_1, \theta_4, \theta_6)
& \sim  - a M_W^4 \cos( \theta_h + 2 \theta_1) \\  
& \\
&  - b (M_W^5/M_{Pl}) \cos (2 \theta_h + \theta_4) 
- c (M_W^7/M_{Pl}^3) \cos (3 \theta_h + \theta_6). 
\end{array}
\end{equation}

Let us now follow the sequence of events in the early universe.
The fields $X_1$, $Y_4$ and $Y_6$ acquire their expectation
values when the temperature is superlarge. This is followed by a
period of inflation that ``irons out" these expectation values,
so that they are virtually constant in the region that is to
become our presently observable universe. Thus, $\theta_1$, $\theta_4$, 
and $\theta_6$ can be treated as constant in space. When the temperature 
falls to a value of order $M_W$, the fields $H_u$ and $H_d$ obtain
expectation values that are of order $M_W$. It is at that point that
the potential for the phases shown in Eq. (8) appears. 
This potential rapidly causes the phase $\theta_h$ to align with
the phase of $\theta_1$ according to the relation
$\theta_h = - 2 \theta_1$. Note that because of the fact that
$\langle X_1 \rangle \gg \langle H_{u,d} \rangle$, it is $\theta_h$
that adjusts, while at this point $\theta_1$ remains virtually
constant in time as well as space. The field corresponding to $\theta_h$
has mass of order $M_W$, and so its oscillations about its minimum should 
should damp rapidly. 

Substituting the relation $\theta_h = - 2 \theta_1$ into the
last two terms in Eq. (8) one gets

\begin{equation}
\begin{array}{cl}
V(\theta_1, \theta_4, \theta_6) & \sim  - b (M_W^5/M_{Pl}) \cos 
( - 4 \theta_1 + \theta_4)
\\  & \\
& - c (M_W^7/M_{Pl}^3) \cos (- 6 \theta_1 + \theta_6). 
\end{array}
\end{equation}

\noindent
Denote the field corresponding to $\theta_1$ by $a_1$. Since
$a_1 = f \theta_1$ it has a mass-squared
$m_{a_1}^2 \sim b (M_W^5/M_{Pl})/f^2 \sim a b (M_W^4/ M_{Pl}^2)$.
This phase will begin coherent oscillations when the age of the
universe is of order $m_{a_1}^{-1}$, or when
$T = T_1 \sim (a b/ g)^{1/4} M_W$, where $g \sim 10^2$ is the number of
massless degrees of freedom when $T = T_1$.  

These oscillations will damp due to the expansion of the universe,
so that by now the phase $\theta_1$ is essentially at the minimum
of the potential given by the first term in Eq. (9), i.e. $\theta_1
= N\pi/2 + \theta_4/4$. The second term in Eq. (9) lifts the degeneracy of 
these minima and leads to an effective cosmological constant
that is of order $c M_W^7/M_{Pl}^3$. 
Demanding that this give the observed
$\Lambda$ of $(2 \times 10^{-3} {\rm eV})^4$, gives 
$M_W \sim c^{-1/7} 30$ GeV. 

In any model where the cosmological constant is caused
by being in a false vacuum, it will eventually
go to zero because of vacuum tunnelling, but this will take a time
much longer than the present age of the universe \cite{gc}.
In this particular model, however, the effective cosmological
constant will eventually go to zero because of the classical evolution
of the fields. After minimizing the first term in Eq. (9), the last
term will depend on the (gauge invariant) phase $\theta \equiv
\theta_6 - 3 \theta_4/2$. Call the properly normalized field 
corresponding to this phase $\overline{a} = \langle Y \rangle \theta$. 
($\langle Y \rangle$ is a linear combination of $\langle Y_4 \rangle$
and $\langle Y_6 \rangle$.) When the age of the universe is
$t \sim m_{\overline{a}}^{-1}$, the field $\overline{a}$ will commence
damped oscillations about the minimum of the last term in Eq. (9).
As it approaches this minimum, the effective cosmological constant
will disappear. Since the coefficient of the last term in Eq. (9)
is, by assumption, of order $\Lambda$, the mass of $\overline{a}$
is given by $m_{\overline{a}} \sim \sqrt{\Lambda}/\langle Y \rangle$.
If $\langle Y \rangle \gg M_{Pl}$, then it is clear that the
oscillations of $\overline{a}$ will not begin until a time much longer than
the present age of the universe. Thus, at the present epoch $\Lambda$
is effectively constant in time.

The reason that there is in this model a residual dynamical phase
upon which $\Lambda$ depends is that the number of terms in Eq. (7) 
happens to be the same as the number of physical phases of the fields.
There is nothing that prevents the construction of models of the
same type, but in which the number of terms in $V$ is sufficient to
give a fixed $\Lambda$, as in the model of Section 2. In such a
model the effective $\Lambda$ would persist 
until our part of the universe tunnelled to the true vacuum. 

Now let us see what the coherent oscillations of $a_1$ 
contribute to the energy density of the universe now.
Using the fact that $(\rho_{a_1}/\rho_B)_{now} =
m_p^{-1}(\rho_{a_1}/n_B)_{now} \sim m_p^{-1} (\rho_{a_1}/n_B)_{T_1}$, 
one has that 

\begin{equation}
\begin{array}{ccl}
(\rho_{a_1}/\rho_B)_{now} & \sim & (b M_W^5/M_{Pl})/(g \eta_B m_p T_1^3) \\
& & \\
& \sim & (g^{-1/4} b^{1/4} a^{-3/4}) \eta_B^{-1} \left( 
\frac{M_W^2}{m_p M_{Pl}} \right) \sim 10^{-5} (b^{1/4} a^{-3/4}).
\end{array}
\end{equation}

\noindent
where $\eta_B \sim 10^{-10}$ is the baryon-to-entropy ratio of the
universe, and we have used here $M_W \sim 10^2$ GeV. If $a
\sim 10^{-8}$ (corresponding to $f \equiv |\langle X_1| \rangle
\sim M_{GUT}$) the energy in the coherent oscillations of the $a_1$ 
has the right magnitude to be the dark matter. 

There is a technical point that should be mentioned that concerns the
naturalness of the hierarchy between the
Weak and Planck scales. A simple way to fix the magnitude of the
field $Y_4$ would be to introduce a field $Y_{-4}$ conjugate to 
it and terms in the superpotential of the form $(Y_4 Y_{-4} - M^2)Z_0$,
where $M \sim M_{Pl}$. However, this would allow the term
$H_u H_d Y_6 Y_{-4}/M_{Pl}$, which would give a Planck-scale
contribution to the $\mu$ parameter and destroy the hierarchy of scales.
However, there are many ways to avoid this problem. For example, there
might be no conjugate field for $Y_4$ but one for $Y_6$, and terms of the form:
$(Y_6 Y_{-6} - M^2)Z_0 + (Y_4^3/M_{Pl} - Y_6^2)Z_{-12}$.

\section{Inflation and alternatives}

In the scenarios of 
Sections 2 and 3 inflation was invoked to explain how the observable
universe ended up in a false vacuum. However, there are non-trivial
constraints that apply to any realistic model of inflation, among which
is that the spectrum of density fluctuations 
produced by inflation must be consistent with observational limits.

For the axion model in Section 2, the phase of 
$\phi$ will take on variations within the observable horizon  
that are of order $H/f_a$, where $H$ is the expansion rate 
during inflation. $H$ must be smaller than $f_a$,
otherwise the inflation-induced fluctuations would simply
randomize the phase of $\phi$,
and the whole scenario would fall apart as the universe would
not end up in the false vacuum. A much stronger constraint on $H$
comes from the effect of fluctuations on 
the mass density of axions $\rho_a$. The density of axions at 
the end of the QCD 
transition is of order $m_a n_a$, where $n_a = \theta_i^2 f_a^2 H_i$ and 
here $H_i$ is the expansion parameter at $T \sim {\rm few} \; \Lambda_{QCD}$ 
when the axion field becomes dynamic. The density depends on the initial 
alignment angle $\theta_i$. Fluctuations in $\theta_i$ translate into 
fluctuations in the axion density \cite{st}. 
In the current context that implies 
$\delta\rho_a/\rho_a = 2 \delta \theta_i/\theta_i 
\sim \frac{H}{f_a\theta_i}$. If axions make up the dark matter,
then $f_a \sim 10^{12}$ GeV, and $\delta \rho/\rho \sim \delta \rho_a/\rho_a$.
Taking $\theta_i \sim 1$, and requiring that $\delta \rho/\rho$
be less than about $10^{-4}$, implies that $H$ must be less than about
$10^8$ GeV. In fact, this limit on $H$ 
applies even if $f_a$ is smaller than $10^{12}$ GeV and
axions are not the dark matter. The point is that
$\delta \rho_a/\rho_a$ scales as $f_a^{-1}$ and 
$\rho_a$ itself scales as $f_a$, implying 
that $\delta \rho_a$ is independent of $f_a$. 

Turning to the models of Section 3, one sees that fluctuations in the
phase $\theta_1$ will introduce fluctuations in the density of matter
$\delta \rho_{a_1}$ similar to the fluctuations $\delta \rho_a$ in 
axion models.
These will be of order $\delta \rho_{a_1} \sim \rho_{a_1} (H/f)$. If
$\rho_{a_1}$ provides the dark matter, the $H$ must be less than
$10^{-4} f \sim 10^7$ GeV.
Similarly, inflation will produce fluctuations
in the effective cosmological constant that we will call $\delta
\rho_{\Lambda}$. Generically, $\delta \rho_{\Lambda} \sim \rho_{\Lambda}
(H/\langle Y \rangle)$. 
From the fact that $\rho_{\Lambda} \sim \rho_{matter}$ when
$z \sim 2$, it follows that $\rho_{\Lambda}$ was only about $10^{-8} 
\rho_{matter}$ at recombination. Thus, even if $\delta \rho_{\Lambda}/
\rho_{\Lambda}$ was of order unity, it would have had a negligible 
effect on the cosmic background radiation at the time of last scattering.
On the other hand, one has to worry about the effects of fluctuations
in $\rho_{\Lambda}$ on the later propagation of the cosmic background
radiation. The magnitude of such effects 
depends on the scale as well as magnitude of the fluctuations. 
Inflation-induced fluctuations in $\rho_{\Lambda}$ would 
cause time-dependent variations in the equation of state with 
roughly equal amplitude on all scales, and therefore the strongest microwave 
background constraint probably comes from COBE observations on large 
angular scales. When these large scales enter the horizon 
$\rho_{\Lambda}/\rho \sim 1$, so that one would have the constraint
$\delta \rho_{\Lambda}/\rho_{\Lambda} < 10^{-5}$. In the 
context of the models in Section 3, this would mean that
$H$ during inflation would have to be less than $10^{14}$~GeV for 
$\langle Y \rangle = M_{Pl}$.

Given the constraints that exist on inflationary models,
it is worth asking whether some other way, not involving inflation, 
might be found to explain how the universe ended up in a false vacuum
There are two ideas that we think are worth discussing
in spite of the fact that we have not found a satisfactory
implementation of them in the context of gravitationally induced
higher-dimension operators.

The first idea is that thermal contributions to the effective potential  
might steer the field $\phi$ into a false vacuum. That is, it may be 
possible that the relative energies of the minima are different at 
high temperatures than at low. The difficulty we have found in realizing 
this possibility
is that the leading finite-temperature effects produced by a certain
term in $V(\phi)$ tend to have the same symmetry as that term.
For example, if there were a term $\phi^4(\phi^{\dag} \phi)/M_{Pl}^2$ in $V$, 
it would lead to a T-dependent correction of the form $T^2 \partial^2/
\partial \phi \partial \phi^\dagger [\phi^4(\phi^{\dag} \phi)/M_{Pl}^2]
\sim T^2 \phi^4/M_{Pl}^2$,
which has the same symmetry $\phi \longrightarrow e^{i N \pi/2} \phi$.

A second idea is that the true vacuum may fail to percolate and end
up losing out to the false vacuum when the domain walls disappear. 
This might happen as follows. Consider a potential 
in which the true vacuum minimum, though deeper, is narrower than the false 
vacuum minimum. For example,  
in a model with a global U(1), suppose the true minimum is at $\theta=0$ 
and the false minimum at $\theta=\pi$, but that the barriers which 
separate the two minima have 
peaks at $\theta=\pm \pi/4$. Then, when $\phi$ starts 
to feel the explicit U(1) breaking $V(\theta)$, there is a probability of 
0.25 that it rolls toward the true vacuum and 0.75 that it rolls 
towards the false vacuum. As a result, the false vacuum phase  
percolates, while the true vacuum phase consists of isolated bubbles. 
Whether those bubbles grow or shrink depends on whether or not 
they exceed some critical size $R_c$ that depends on the surface tension 
of the walls and the pressure difference between the vacua. Generally, 
this length scale is the same as the time scale  
for squeezing out the false vacuum. So, if the true vacuum does not 
percolate, it is the true vacuum that gets squeezed out, not the false 
vacuum. Although this scheme has a simple appeal, we have not been able 
to construct a simple model that naturally has the required properties. 
In the paradigm of Eq. 2, a single term dominates $V(\theta)$ to produce 
several degenerate minima. For a U(1) model it is apparent that these 
minima are equally spaced, the true vacua will be just as common as the 
others, and when it comes time to dynamically choose a vacuum, the true 
vacuum will win out by virtue of pressure differences. We suspect this is 
generally true for more complicated symmetries. From a different 
perspective, a power series in $\cos \theta$ can be constructed which has 
the required shape, but the terms must all be of the same order of 
magnitude. In the context of gravitationally induced higher dimensional 
operators, however, each term in such a power series would come 
suppressed by correspondingly higher powers of $M_{Pl}$ and so it is not 
natural for them to be of the same order of magnitude.


\begin{thebibliography}{999}

\bibitem{sn} S.J. Perlmutter et al. {\it Nature} {\bf 391}, 51 (1998);
S.J. Perlmutter et al. {\it Astrophys. J.} 517 (1999); A.G. Reiss
et al. {\it Astron. J.} {\bf 116}, 1009 (1998).

\bibitem{cosprinc} {\it Gravitation and Cosmology}, S. Weinberg
John Wiley and Sons, New York, 1972), p 407-412.

\bibitem{ccp} S.Weinberg, {\it Rev. Mod. Phys.} {\bf 61}, 1 (1989).

\bibitem{ahkm} N. Arkani-Hamed, L.J. Hall, C. Kolda, and H. Murayama,
{\it Phys. Rev. Lett.} {\bf 85}, 4434 (2000).

\bibitem{ap} {\it The Cosmological Anthropic Principle}, J.D. Barrow
and F.J. Tipler (Oxford Univ. Press, 1986).

\bibitem{weinberg} S. Weinberg, {\it Phys. Rev. Lett.} {\bf 59}, 2607 (1987);
{\it Phys. Rev.} {\bf D61}, 103505 (2000).


\bibitem{wetterich} C. Wetterich, {\it Nucl. Phys.} {\bf B302}, 668 (1988).

\bibitem{peebles} 
P.J.E. Peebles and B. Ratra, {\it Astrophys. J.} {\bf 325},
L17 (1988); B. Ratra and P.J.E. Peebles, {\it Phys. Rev.} 
{\bf D37}, 3406 (1988).
 
\bibitem{quint} Y. Fujii and T. Nishioka, 
{\it Phys. Rev.} {\bf D42}, 361 (1990);
J. Frieman, C. Hill, and R. Watkins, {\it Phys. Rev.}
{\bf D46}, 1226 (1992); E.J. Copeland, A.R. Liddle, and D. Wanda,
{\it Ann. N.Y. Acad. Sci.} {\bf 688}, 647 (1993);
J. Frieman, C. Hill, A. Stebbins, and I. Waga,
{\it Phys. Rev. Lett.} {\bf 75}, 2077 (1995); K. Coble, S. Dodelson,
and J.A. Frieman, {\it Phys. Rev.} {\bf D55}, 1851 (1997);
R.R. Caldwell, R. Dave, and P.J. Steinhardt, {\it Phys. Rev. Lett.}
{\bf 80}, 1582 (1998).

\bibitem{track} P. Ferreira and M. Joyce, {\it Phys. Rev. Lett.}
{\bf 79}, 474 (1997); I. Zlatev, L. Wang, and P.J. Steinhardt,
{\it Phys. Rev. Lett.} {\bf 82}, 896 (1999).

\bibitem{barr} S.M. Barr, {\it Phys. Rev.} {\bf D36}, 1691 (1987).

\bibitem{hw} A. Hebecker and C. Wetterich, {\it Phys. Lett.}
{\bf B497}, 281 (2001).

\bibitem{gc} W. Garretson and E. Carlson, {\it Phys. Lett.} {\bf B315},
232 (1993).

\bibitem{goldberg} H. Goldberg, {\it Phys. Lett.} {\bf B492}, 153 (2000).

\bibitem{pq} R. Peccei and H. Quinn, {\it Phys. Rev. Lett.} {\bf 38},
1440 (1977).

\bibitem{holman} R. Holman, S.D.H. Hsu, T.W. Kephart, E.W. Kolb,

\bibitem{kam} M. Kamionkowski and J. March-Russell, {\it Phys. Lett.}
{\bf B282}, 137 (1992).

\bibitem{bs} S.M. Barr and D. Seckel, {\it Phys. Rev.} {\bf D46}, 539 (1992).
R. Watkins, and L.M. Widrow, {\it Phys. Lett.} {\bf B282}, 132 (1992).

\bibitem{bb} K.S. Babu and S.M. Barr, {\it Phys. Lett.} {\bf B300}, 367 (1993).

\bibitem{ksvz} J.E. Kim, {\it Phys. Rev. Lett.} {\bf 43}, 103 (1979);
M.A. Shifman, A.I. Vainshtein, and V.I. Zakharov, {\it Nucl. Phys.}
{\bf B166}, 493 (1980).

\bibitem{dfsz} M. Dine, W. Fischler, and M. Srednicki, {\it Phys. Lett.}
{\bf B104}, 199 (1981); A.R. Zhitnitsky, {\it Sov. J. Nucl. Phys.}
{\bf 31}, 260 (1980).

\bibitem{st} D. Seckel and M.S. Turner, {\it Phys. Rev.} {\bf D32}, 
3179 (1988).

\end{thebibliography}
\end{document}